\documentclass[10pt,journal]{IEEEtran}

\usepackage[T1]{fontenc}
\usepackage[utf8]{inputenc}
\usepackage{graphicx}
\usepackage{textcomp}
\usepackage{booktabs}
\usepackage{array}
\usepackage[hyphens]{url}
\usepackage[colorlinks=true,citecolor=blue,urlcolor=blue,linkcolor=blue]{hyperref}

\begin{document}

\title{Computing at Sea: Floating and Offshore Data Centres as a Pathway to Sustainable AI Infrastructure}

\author{Cheng Siong Chin, Jianhua Zhang, and M. Venkateshkumar}

\markboth{}{}

\maketitle

\begin{abstract}
The rapid expansion of artificial intelligence is transforming data centres into one of the world's fastest-growing sources of electricity demand. As AI systems scale in size and capability, the physical infrastructure supporting computation is approaching critical limits in energy availability, cooling capacity, land use, freshwater consumption, and carbon management. Conventional land-based data centres are increasingly constrained by urban land competition, grid congestion, environmental pressures, and lengthy permitting processes, raising fundamental questions about where future computing infrastructure can sustainably exist. This article examines floating and offshore data centres as an emerging alternative model for digital infrastructure. By relocating computation to marine environments, offshore systems can exploit the ocean's natural cooling capacity, reduce freshwater dependence, and enable direct integration with offshore renewable energy resources such as wind, wave, and tidal power. Early deployments have demonstrated the potential for significantly improved energy efficiency and operational reliability compared with conventional facilities, while also opening new possibilities for distributed and resilient computing architectures. The article explores how offshore computing may reshape the future relationship between electrification, renewable energy, and large-scale AI infrastructure. It analyses the opportunities and trade-offs associated with marine deployment, including environmental impacts, engineering design challenges, economic feasibility, and regulatory governance. Rather than treating offshore data centres as experimental novelties, the article presents them as part of a broader systems-level transition in how society may power, cool, and sustain the next generation of computational growth.
\end{abstract}

\section{Introduction}

At any given moment, vast numbers of people are streaming content, consulting AI tools, moving money across financial networks, or pushing data to the cloud. Each of these everyday actions draws on an increasingly sprawling global network of hyperscale data centers, industrial-scale facilities that, taken together, consumed somewhere between 300 and 380 TWh of electricity in 2023 \cite{ref1}, a figure roughly on par with the United Kingdom's entire annual power demand. The International Energy Agency projects \cite{ref2} that this appetite will nearly double to around 945 TWh by 2030, growing at approximately 15\% per year from 2024, a pace more than four times that of global electricity consumption \cite{ref2}. In the United States, data centers already account for around 4.4\% of national electricity use, having consumed 176 TWh in 2023, with estimates for 2028 ranging as high as 580 TWh \cite{ref3}.

What makes this trajectory remarkable is its concentration in a single sector over a historically short timeframe. The Electric Power Research Institute has estimated that AI alone accounted for between 10 and 20\% of all data center energy use in 2024 \cite{ref4} and running a single generative AI query demands roughly four to five times the energy of a standard web search \cite{ref5}. Capital has followed suit: global investment in data center infrastructure nearly doubled between 2022 and 2024, reaching half a trillion dollars \cite{ref2}. ABI Research projects \cite{ref6} that worldwide data center power consumption will more than double between 2024 and 2030, from 683 TWh to nearly 1,479 TWh, representing a compound annual growth rate of around 14\%. The trajectory has not gone unnoticed at the highest levels; as OpenAI's chief executive remarked at Davos in early 2025, the industry is on a collision course with an energy crisis \cite{ref5}. Table~\ref{tab:demand} summarizes global and US data centre energy consumption from 2018 to 2024 and projects demand through 2030, illustrating the accelerating trajectory driven sequentially by hyperscale expansion, cloud adoption, and generative AI proliferation \cite{ref1,ref2,ref3}.

\begin{table*}[t]
\centering
\caption{Historical consumption figures for 2018--2024 and projections through 2030 illustrate accelerating energy demand driven sequentially by hyperscale expansion, cloud adoption, and generative AI proliferation. US consumption is shown alongside global totals to reflect the continued concentration of data centre infrastructure in North American markets. Projected ranges for 2028 and 2030 reflect uncertainty across AI adoption trajectories, edge computing growth, and data sovereignty-driven infrastructure regionalisation.}
\label{tab:demand}
\begin{tabular}{p{1.6cm}p{2.6cm}p{2.4cm}p{2.6cm}p{5.4cm}}
\toprule
\textbf{Year} & \textbf{Global DC Consumption} & \textbf{\% Global Electricity} & \textbf{US DC Consumption} & \textbf{Key Driver} \\
\midrule
2018 & $\sim$200 TWh & $\sim$1.0\% & $\sim$73 TWh & Hyperscale expansion \\
2023 & 300--380 TWh & $\sim$1.4\% & 176 TWh & Cloud and early AI adoption \\
2024 & $\sim$415 TWh & $\sim$1.5\% & $\sim$200 TWh & Generative AI surge \\
2028 (proj.) & 683--960 TWh & $\sim$2.5\% & 325--580 TWh & AI acceleration; edge growth \\
2030 (proj.) & 945--1{,}479 TWh & 3--4\% & $>$700 TWh & AI, edge, data sovereignty \\
\bottomrule
\end{tabular}
\end{table*}

These pressures do not exist in isolation; they are converging into a compounding resource crisis. Cooling alone typically absorbs 30--40\% of a conventional data center's electricity consumption \cite{ref7}, and the water demands are equally striking. A mid-scale facility running at 15 MW draws water at a rate comparable to three hospitals or a pair of 18-hole golf courses \cite{ref8}. The real-world consequences of this are already visible: a data center cluster in West Des Moines, Iowa, used to support the training of GPT-4 drew down roughly 6\% of the local water district's monthly supply in a single month \cite{ref5}. Over the course of large language model development, Google and Microsoft reported year-on-year increases in water consumption of 20\% and 34\%, respectively \cite{ref5}. In land-scarce, power-constrained markets the situation has already forced regulatory intervention, for example, Singapore placed a moratorium on new land-based data center approvals between 2019 and 2022, unable to accommodate further demand within its existing infrastructure \cite{ref9}. It is against this backdrop that a growing number of engineers and technology firms are pursuing a more radical proposition: relocating data centers to the ocean. Floating and offshore facilities, whether mounted on surface platforms or deployed on the seafloor, represent a genuinely different way of operating this infrastructure. The sea offers a thermally stable, naturally circulating heat sink of effectively unlimited scale, enabling passive or near-passive cooling with no freshwater draw whatsoever. The marine environment also places facilities within reach of substantial offshore renewable resources, including wind, wave, tidal, and ocean thermal energy. Early evidence from real deployments is encouraging: recent commercial installations in China have reported PUE figures below 1.3 \cite{ref10}, while Microsoft's Project Natick, a sealed subsea module that operated on the seafloor off Scotland's Orkney Islands for 2 years, recorded server failure rates lower than comparable land-based systems, all without consuming a single litre of freshwater \cite{ref11}.

\section{Floating Data Center Architecture}

\begin{figure*}[t]
\centering
\includegraphics[width=0.72\textwidth]{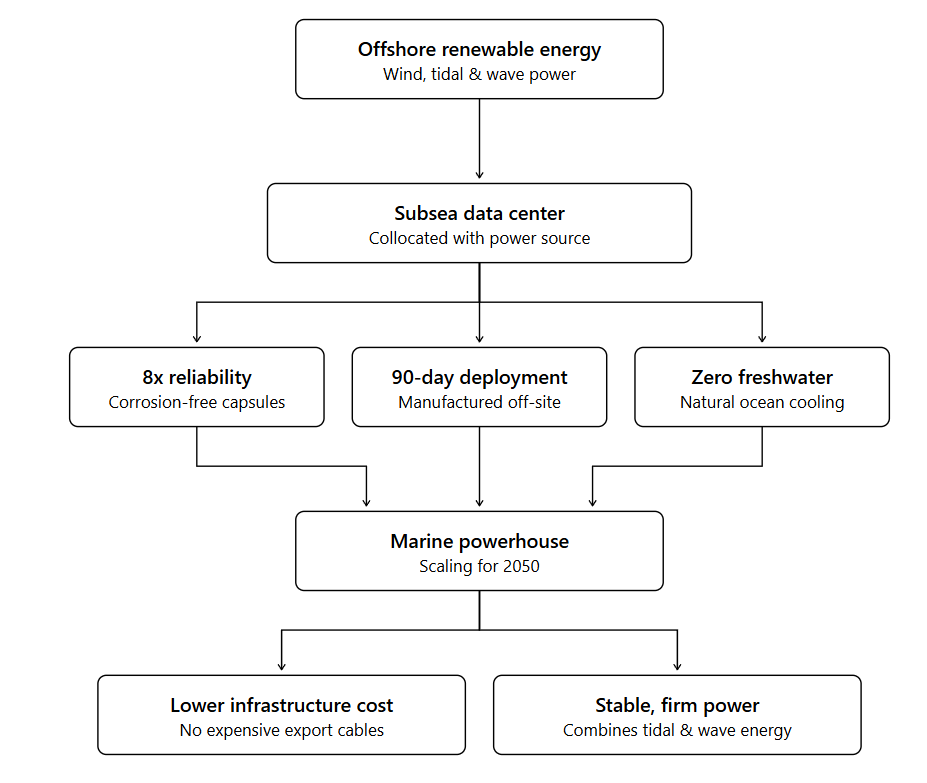}
\caption{Flow of the subsea data center concept: offshore renewable energy (wind, tidal, and wave) feeds a collocated subsea data center, which delivers three core advantages (8x reliability, 90-day deployment, zero freshwater use). These combine to enable the ``marine powerhouse'' vision scaled for 2050, producing lower infrastructure cost and a stable, firm power supply.}
\label{fig:concept}
\end{figure*}

Fig.~\ref{fig:concept} traces the logic of the subsea data centre concept from power source to outcome. Offshore renewable energy, drawn from wind, tidal, and wave sources, feeds a subsea data centre collocated at the generation site. This collocation yields three core advantages: an eightfold gain in reliability from corrosion-free capsule design, a 90-day deployment window enabled by off-site manufacturing, and zero freshwater consumption through natural ocean cooling. Together, these advantages support the ``marine powerhouse'' concept, a scaled deployment model envisioned for 2050, which in turn delivers two outcomes: lower infrastructure cost, since expensive export cables become unnecessary, and a stable, firm power supply from the combination of tidal and wave energy. China leads commercialisation in this domain. Beijing Highlander Digital Technology's Hainan facility \cite{ref10} deployed an initial submerged module in 2023 at 30--40 m depth, connected to shore via composite submarine cables transmitting both power and data. The facility substantially reduces freshwater consumption relative to land-based equivalents. Additional multi-module clusters are proposed across the Yangtze River Delta and Pearl River Delta \cite{ref12}.

The most technically ambitious approach places sealed, pressure-resistant capsules on the ocean floor. Microsoft's Project Natick \cite{ref11} remains the definitive documented example. The Phase 2 capsule, 12.2 m in length housed 864 Azure servers across 12 racks and 27.6 petabytes of storage, operating at 35.6 m depth for 26 months via a single subsea power and data cable. Cooling was entirely passive, with seawater circulating through an external hull heat exchanger. A nitrogen interior atmosphere eliminated corrosion and reduced humidity to near zero, contributing to server failure rates approximately eightfold lower than comparable land-based facilities, with significant implications for reliability engineering and lifecycle carbon accounting. Containerised systems adapt standard ISO shipping containers for deployment on semi-submersible barges, repurposed offshore platforms, or purpose-built pontoons. A module housing 20 server racks can be assembled and stress-tested onshore, then deployed within weeks, contrasting sharply with the 18--36 months required for greenfield hyperscale construction. Capacity is added or removed one module at a time, decoupling infrastructure investment from demand forecasting and enabling incremental scalability, a strategic necessity in environments where AI workload volumes shift rapidly over short periods.

Co-designing energy generation and computing infrastructure from the outset unlocks compelling integration advantages. Offshore wind turbines mounted on or near the platform minimise transmission distances and associated losses. Wave energy converters moored alongside provide a complementary generation source that partially offsets wind intermittency. In tropical and subtropical waters, ocean thermal energy conversion (OTEC) exploits the temperature gradient between warm surface water and cold deep water to simultaneously generate electricity and deliver chilled water to server infrastructure, a unified thermodynamic system with no land-based equivalent. Collectively, these co-location strategies eliminate the transmission losses typical of onshore alternating current networks, moving offshore platforms toward genuinely self-contained energy islands largely decoupled from onshore grid constraints.

\section{Energy Consumption Comparison}

The fundamental energy advantage of offshore data centers derives from the thermodynamic properties of seawater as a cooling medium and the consequent elimination of mechanical refrigeration. Quantifying this advantage requires examining four key metrics: Power Usage Effectiveness (PUE), Water Usage Effectiveness (WUE), cooling energy fraction, and infrastructure energy overhead. PUE is defined as the ratio of total data center facility energy consumption to IT equipment energy consumption. A value of 1.0 represents theoretical perfection; every watt consumed serves useful computation. The global average PUE for all data centers was approximately 1.58 in 2023, reflecting the continued operation of legacy facilities with inefficient air-conditioning infrastructure \cite{ref13}. Modern hyperscale land-based facilities such as Google's global fleet average was 1.09 in 2025 \cite{ref14}. Offshore seawater-cooled facilities demonstrate a measurable further improvement. Surface-floating platforms with heat exchangers have achieved sustained PUE values of 1.11--1.15 in commercial operation \cite{ref15}. Microsoft's Project Natick, with entirely passive seawater cooling in a nitrogen-filled pressure vessel, achieved an estimated PUE of approximately 1.07 \cite{ref11}. Table~\ref{tab:pue} summarizes PUE values \cite{ref1,ref11} across data center generations, illustrating the progressive improvement from legacy air-cooled to passive underwater seawater cooling.

\begin{table*}[t]
\centering
\caption{Power Usage Effectiveness across data centre generations. PUE and cooling energy decline progressively from legacy air-cooled facilities to offshore seawater-cooled systems. Microsoft's Project Natick achieved a PUE near 1.07 using passive seawater cooling with zero freshwater consumption.}
\label{tab:pue}
\begin{tabular}{p{3.0cm}p{1.6cm}p{1.4cm}p{1.7cm}p{5.9cm}}
\toprule
\textbf{Data Center Type} & \textbf{PUE Range} & \textbf{Typical PUE} & \textbf{Cooling \% of Energy} & \textbf{Cooling Approach} \\
\midrule
Traditional Land (Legacy) & 1.8--2.5 & 2.0 & 35--40\% & Air-cooled chillers, cooling towers \\
Modern Hyperscale (Land-Based) & 1.1--1.4 & 1.2 & 15--25\% & Free air and evaporative cooling \\
Best-in-Class Hyperscale (Land) & 1.03--1.12 & 1.10 & 8--15\% & Liquid cooling or cold-climate free air \\
Offshore Surface Platform & 1.05--1.15 & 1.10 & 5--10\% & Seawater heat exchange loop \\
Microsoft Project Natick (Seafloor) & $\sim$1.07 & 1.07 & $\sim$7\% & Passive seawater hull heat exchange \\
Deep Underwater (Future Design) & 1.03--1.07 & 1.05 & 3--5\% & Passive deep-seawater thermosiphon \\
\bottomrule
\end{tabular}
\end{table*}

\subsection*{Cooling Energy Fraction and Freshwater Elimination}

In a traditional air-cooled land-based data center, mechanical refrigeration (such as chillers, cooling towers, and computer room air conditioning (CRAC) units) accounts for 30\% of total facility energy consumption \cite{ref7}. Transitioning to free-air cooling in temperate climates reduces this to 10--15\%. Seawater-cooled offshore facilities reduce cooling's share to 3--8\%, with the residual representing seawater pump energy and ancillary loads. The passive underwater capsule approaches the physical floor of approximately 2--3\%. Water usage is even more dramatically affected. A conventional cooling-tower-based facility records a WUE of 1.8--2.5 L/kWh; a 1 MW facility consumes approximately 25.5 million liters of fresh water per year \cite{ref7}. Google consumed 15.8 billion liters and Microsoft consumed 3.6 billion liters globally in FY2018 \cite{ref7}. For offshore seawater-cooled facilities, the freshwater WUE is essentially zero: seawater circulates through the heat exchanger, but no freshwater is consumed and no water is evaporated to the atmosphere.

Cooling is not the sole source of energy overhead in conventional data centres. Land-based facilities carry persistent background loads such as lighting, security, fire suppression, and administrative services, that consume energy independently of server utilisation. Offshore platforms, designed around computation rather than human occupancy, shed much of this burden by default. With automation handling routine operations and maintenance personnel visiting periodically rather than staffing continuously, non-IT energy draw shrinks considerably. This operational philosophy mirrors that of unmanned offshore oil and gas platforms, where decades of engineering have produced reliable, remotely supervised infrastructure without permanent crews. Applied to data centres, the same logic yields direct PUE improvement, not through advances in server technology or cooling chemistry, but by eliminating the incidental energy demands that accumulate whenever a facility must sustain human presence alongside compute workloads.

Connecting offshore platforms to onshore grid and network infrastructure introduces transmission overhead, though the penalty is more modest than commonly assumed. Beyond approximately 50--60 km, high-voltage direct current (HVDC) technology is the established choice for submarine power delivery, as reactive power losses limit submarine AC cables to practical lengths of around 50 km. For nearshore deployments of 50--200 km, the distance range relevant to current and planned offshore installations, total system losses remain comparable to equivalent onshore AC transmission. Network latency is similarly bounded by well-understood physics: single-mode optical fibre propagates light at approximately two-thirds the speed of light in vacuum, yielding a round-trip propagation penalty of approximately 0.5--1.0 ms for a facility located 50--100 km offshore. AI model training workloads are latency-insensitive, tolerating inter-node delays of up to 100 ms across jobs extending days to weeks, rendering geographic proximity to end-users irrelevant to performance. Workloads genuinely requiring sub-millisecond response, real-time application inference, high-frequency financial transactions, represent a specialised segment requiring proximity-optimised placement irrespective of whether the hosting facility is offshore or land-based.

\section{Renewable Energy Integration}

\begin{figure*}[t]
\centering
\includegraphics[width=0.72\textwidth]{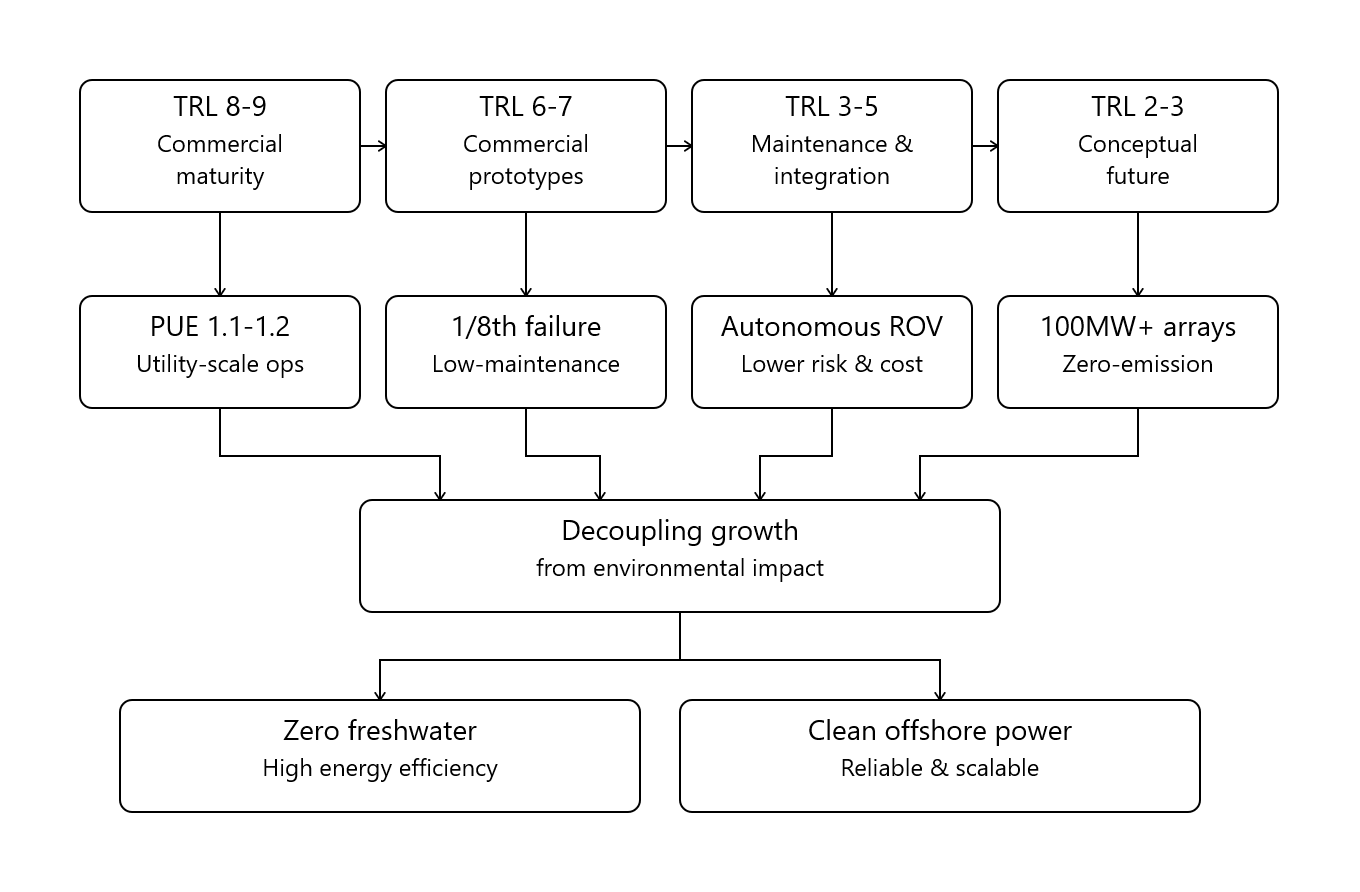}
\caption{Technology readiness roadmap for subsea data centers, progressing from TRL 8--9 (commercial maturity, e.g. seawater cooling and surface platforms) through TRL 6--7 (commercial-scale prototypes such as sealed nitrogen-filled capsules) and TRL 3--5 (autonomous ROV-based maintenance and integration) to TRL 2--3 (conceptual large-scale compute-archipelago futures). Each stage's key feature feeds into the goal of decoupling computing growth from environmental impact, culminating in zero-freshwater, high-efficiency operation and clean, scalable offshore power integration.}
\label{fig:trl}
\end{figure*}

The energy case for offshore deployment is structurally distinct from any land-based equivalent, as illustrated by the TRL roadmap in Fig.~\ref{fig:trl}. Unlike onshore facilities, where renewable procurement is complicated by competing land uses, permitting delays, and transmission losses, offshore infrastructure can draw on diverse co-located marine energy resources to achieve near-zero-carbon operation with fewer systemic barriers. Commercial offshore wind turbines have reached rated capacities of 15--16 MW, with 17--20 MW variants under active development and announced designs from Mingyang Smart Energy projecting up to 22 MW per unit. A representative five-turbine, 75 MW array would yield 30--41 MW of time-averaged output at capacity factors of 40--55\%, sufficient to sustain a 30--35 MW IT load when supplemented by short-duration battery storage.

In tropical and subtropical latitudes (approximately 20\textdegree N--20\textdegree S), OTEC offers a particularly compelling integration pathway, corresponding to the TRL 3--5 integration frontier identified in Fig.~\ref{fig:trl}. Critically, the cold deep-water pumped as a process byproduct provides an essentially free cooling medium for co-located IT equipment, unifying power generation and thermal management in a single marine system with no land-based analogue. Wave energy converters, with capacity factors of 20--35\%, provide temporal complementarity to wind generation by sustaining output for several hours after wind speeds drop, thereby improving hybrid microgrid stability. Tidal stream generators, achieving capacity factors of 35--45\%, are well-suited to strong tidal regimes; the European Marine Energy Centre in Orkney exemplifies this potential and was a primary factor in Microsoft's site selection for Project Natick, where the subsea capsule drew power exclusively from tidal and wind generation \cite{ref11}.

An optimal offshore data centre power architecture integrates two or more generation sources with battery storage and a shore-cable backup, coordinated by real-time energy management software that maintains power quality within IT infrastructure tolerances. AI-driven workload scheduling adds further gains in grid responsiveness. As Fig.~\ref{fig:trl} shows, this convergence is what ultimately realises the TRL 2--3 compute archipelago vision: offshore data centres take on an active role within the broader architecture of offshore energy systems, absorbing generation variability, stabilising microgrids, and strengthening the commercial case for marine renewable deployment at scale.

\section{Environmental Impact}

Operational carbon intensity is governed primarily by grid carbon content and total energy consumption. A further, underexamined lifecycle benefit arises from improved server reliability in submerged deployments. Project Natick demonstrated an eightfold reduction in failure rates relative to land-based baselines \cite{ref11}, implying commensurate reductions in hardware manufacturing over the facility lifetime. Given embodied carbon estimates of 300--800 kg CO$_2$-equivalent per server, systematic reliability gains yield material lifecycle savings not yet captured by conventional carbon accounting frameworks.

However, offshore deployment raises substantive environmental considerations requiring rigorous pre-installation impact assessment. Seafloor anchoring disturbs benthic communities during installation; however, experience with offshore wind foundations demonstrates that hard substrates are rapidly colonised, effectively functioning as artificial reefs that can enhance local biodiversity. Thermal effluent from seawater cooling loops, typically 3--5\textdegree C above ambient, is comparable to existing industrial marine discharges and represents a negligible perturbation relative to natural ocean heat fluxes at open-water sites. Deep-water intake systems require screening infrastructure and constrained flow velocities to minimise entrainment of larval and juvenile fish, for which empirical thresholds are well established. Most maritime jurisdictions mandate comprehensive Environmental Impact Assessments, providing structured mechanisms for ecological risk management and enforceable monitoring obligations throughout facility operation.

A 100 MW data centre dissipates approximately 85--90 MW of waste heat to its cooling medium. In conventional land-based facilities, this contributes to urban heat island effects, an externality under growing regulatory scrutiny. In offshore seawater-cooled deployments, heat is discharged directly to the ocean; at current scales, thermal contributions to open-water sites remain negligible relative to natural heat fluxes. Facilities sited in enclosed or semi-enclosed water bodies present a materially different risk profile, where restricted circulation can generate localised temperature anomalies of ecological significance. In such settings, rigorous thermal dispersion modelling is warranted, particularly where thermally sensitive ecosystems, including coral reefs and cold-water fish spawning grounds, may respond disproportionately to even modest sustained temperature increases.

\section{Economic Considerations}

Offshore data centre capital expenditure diverges substantially from land-based facilities. Urban land acquisition, representing 15--25\% of total development cost at prime co-location sites, is replaced by platform procurement and subsea cable installation. Purpose-built offshore platforms are estimated at USD 15--25 million per 10 MW of installed capacity, broadly comparable to land-based construction once real estate premiums are included. Modular containerised approaches reduce unit costs further; a 1 MW floating barge module is estimated at USD 3 million prior to cable infrastructure. The principal capital cost distinctive to offshore deployment is submarine cabling. For clustered multi-operator deployments sharing common cable infrastructure, this burden can be apportioned across participants, substantially improving per-operator economics.

Offshore operational expenditure advantages become increasingly compelling over medium and long-time horizons. Water procurement, treatment, and disposal costs, significant in water-stressed jurisdictions, are effectively eliminated. Reduced server failure rates further decrease hardware replacement frequency over the facility lifetime. In aggregate, operational savings over a ten-year horizon in energy- and water-cost-intensive markets, including Singapore, California, Arizona, and Japan, are sufficient to offset the initial capital premium of offshore platforms and submarine cables, yielding a favourable total cost of ownership trajectory relative to land-based alternatives.

The principal operational expenditure disadvantage of offshore facilities is maintenance accessibility. Surface platforms require vessel-based servicing, while sealed submerged capsules demand ROV intervention and potentially full retrieval for significant hardware failures. Project Natick demonstrated continuous operation for at least 26 months without maintenance intervention \cite{ref11}, and commercial designs now target five-year intervals. The industry is converging on three complementary strategies, namely: subsystem redundancy for critical components; AI-assisted remote diagnostics via continuous sensor telemetry; and modular swap-out architectures enabling failed subsystems to be isolated and replaced without disrupting adjacent compute infrastructure.

Land-based hyperscale data centre construction typically requires 18--36 months from groundbreaking to full capacity, constrained by planning approvals, grid connection queues exceeding five years in several major markets, and labour availability. Modular offshore systems, prefabricated in shipyard environments and fully load-tested prior to deployment, can be commissioned within 6--12 months of contract award. This accelerated trajectory carries strategic significance as hyperscale operators face mounting pressure to provision AI compute capacity at a pace increasingly misaligned with conventional land-based construction timelines.

\section{Industry Case Studies}

Project Natick remains the most comprehensively documented offshore data centre experiment to date. Phase 1 deployed off California in 2015, established proof-of-concept feasibility at single rack scale. Phase 2, deployed at the European Marine Energy Centre off Orkney in June 2018, scaled to 12 racks housing 864 Azure servers and 27.6 petabytes of storage, powered entirely by Orkney's renewable grid \cite{ref11}. After 26 months of uninterrupted seafloor operation at 35 m depth, the capsule was retrieved in July 2020. Server failure rates were eightfold lower than land-based equivalents, freshwater consumption was zero, and no unplanned maintenance was required. The pressurised nitrogen atmosphere eliminated corrosion, thermal stability reduced solder fatigue, and the absence of human activity removed physical disturbance as a failure mode. Microsoft is now evaluating commercial deployment pathways, particularly for edge computing, citing that over half the global population lives within 200 km of a coastline.

China has achieved the most rapid transition from concept to commercial operation in this domain. Hainan Highlander Digital Technology's South China Sea platform, the world's first operational offshore floating data centre entered service in 2023, accommodating approximately 400 server racks at a power usage effectiveness below 1.15 \cite{ref15}. Seawater drawn from 20 m depth circulates through plate heat exchangers, returning at 3--4\textdegree C above ambient. Multiple additional deployments are advancing along Guangdong's Pearl River Delta coastline, driven by regional ambitions to establish an AI compute hub and by acute land and power constraints in China's coastal technology clusters. The Chinese government has formally designated offshore data infrastructure as a strategic priority within its carbon neutrality framework, accelerating commercial deployment at a scale unmatched elsewhere.

Though not a floating facility, Google's Hamina data centre represents the most operationally mature demonstration of seawater cooling at hyperscale. Established in 2011 within a repurposed paper mill on the Gulf of Finland, the facility circulates cold gulf water through dedicated heat exchangers, achieving zero freshwater consumption across more than a decade of continuous operation. A USD 670 million capacity expansion in 2022 underscores its strategic importance, and a reported power usage effectiveness of 1.09 places it among the most thermally efficient hyperscale installations globally. Hamina provides compelling longitudinal evidence that seawater cooling is a reliable, scalable, and commercially sustainable thermal management approach, with a track record sufficient to inform investment decisions at the frontier of offshore deployment.

US-based venture Subsea Cloud is advancing commercial underwater data centre capsules for deployment in Puget Sound and the North Sea. Each aluminium pressure vessel, approximately 5 m in diameter and 14 m in length, houses eight server racks in a humidity-controlled atmosphere. Unlike Project Natick's sealed nitrogen architecture, the design incorporates a diver-accessible maintenance hatch enabling component-level replacement without full capsule retrieval, materially reducing servicing cost and logistical burden. Initial commercial deployments are targeted for 2025--2026, focusing on edge computing and content delivery for coastal metropolitan markets, a segment where offshore underwater data centres offer structurally advantaged latency and competitive cost positioning relative to land-based alternatives.

\section{Challenges and Future Directions}

Project Natick demonstrated encouraging reliability, yet sealed submerged facilities remain fundamentally constrained by maintenance inaccessibility. Planned service intervals target five years, but unplanned failures demand full capsule retrieval, requiring specialist vessels and ROVs over several days. Next-generation designs must therefore incorporate greater subsystem redundancy, robotic modular replacement, and machine learning-driven predictive diagnostics informed by continuous sensor telemetry, including large language model-based anomaly interpretation pipelines capable of synthesising multi-stream operational data into actionable maintenance decisions without requiring direct human intervention at the deployment site. Surface platforms offer easier access but still require marine vessel support and remain vulnerable to adverse weather, a limitation that commercial specifications must systematically address.

Offshore data centres occupy an unresolved regulatory intersection of maritime law, telecommunications governance, data sovereignty, and environmental protection. Within 200 nautical-mile exclusive economic zones, coastal state jurisdiction over IT infrastructure in the water column remains legally untested in most jurisdictions. Instruments such as the EU General Data Protection Regulation and Singapore's Personal Data Protection Act may impose territorial data residency requirements. Operators must therefore route regulated data to compliant jurisdictions, reserving offshore facilities for latency-insensitive workloads, AI model training, rendering, archival storage, and scientific computing, unconstrained by sovereignty obligations.

Offshore data centre integration with mainland grids requires bidirectional submarine cables, power-quality management, and ancillary service market participation. As controllable large continuous loads, data centres represent a uniquely valuable demand-flexibility resource. Offshore facilities with direct wind generation access are ideally placed to amplify these benefits, creating a virtuous cycle of renewable integration, cost reduction, and grid stabilisation. AI is already transforming onshore data centre energy management. Offshore deployments offer analogous and amplified gains: optimising seawater pump scheduling against real-time ocean temperature profiles; forecasting renewable generation from meteorological and oceanographic models; dynamically allocating training workloads to match generation availability; and coordinating with offshore wind systems to maximise renewable utilisation. Long-term scalability depends on advances across several enabling technologies: terabit-scale data and power co-transmission via next-generation submarine cables; autonomous underwater robot platforms for maintenance and modular replacement; pressure-tolerant liquid immersion cooling; and purpose-built autonomous service vessels. Looking further ahead, distributed offshore computing arrays, compute archipelagos integrated with utility-scale wind installations could deliver resilient, near-zero-carbon computational services to coastal megacities. For Southeast Asia, where land scarcity is acute, electricity grids face rapid electrification stress, and offshore renewable resources remain largely untapped, this architecture is both technically realisable and economically compelling within the current decade, positioning the region as a natural proving ground for marine digital infrastructure.

\section{Conclusion}

The rapid growth of AI infrastructure is exceeding assumptions in current data centre planning, grid expansion, and siting policy. While land-based systems are increasingly constrained by physical, water, and jurisdictional limits, offshore and floating data centres offer a structurally different paradigm by shifting computation into environments with abundant thermal capacity and renewable energy access. Evidence is now sufficient to reassess their status. Demonstrations such as Project Natick, along with commercial offshore deployments and long-term seawater cooling operations, indicate that technical viability is largely established. The remaining barriers are not engineering-centric but coordination-based, spanning regulation, maritime governance, grid interconnection, and environmental oversight. Addressing these gaps requires joint action across policymakers, regulators, utilities, and the research community. Key priorities include lifecycle carbon and water accounting, assessment of offshore renewable integration benefits, and ecological validation of marine thermal use. The combined pressures of accelerating AI energy demand and global decarbonisation make offshore computing a timely infrastructure option. However, realising its potential depends equally on developing an interdisciplinary workforce spanning marine engineering, subsea systems, power integration, and data centre operations, capabilities that are currently fragmented across separate domains. Ultimately, offshore data centres highlight an underused oceanic thermal and energy resource that may become central to future digital infrastructure, provided institutional, regulatory, and educational frameworks evolve in step with technological progress.

\end{document}